\documentclass{appolb}%
\usepackage{epsfig}
\usepackage{amsmath}
\usepackage{amsfonts}
\usepackage{amssymb}
\usepackage{graphicx}%

\begin{document}

\title{Saturation and Scaling of Multiplicity, Mean $p_{\text{T}}$ and $p_{\text{T}}$
Distributions from $200~\mathrm{GeV}\leq\sqrt{s}\leq7~\mathrm{TeV} $ -- \\Addendum}
\author{Larry McLerran \address{BNL and Riken Brookhaven Center, Upton, NY }
\and Michal Praszalowicz \address{
M. Smoluchowski Institute of Physics, Jagellonian University, Reymonta 4,
30-059 Krakow, Poland} }
\maketitle

\begin{abstract}
In the previous paper we have argued that the LHC data on multiplicity,
average transverse momentum, and charged particle transverse momentum
distributions are well described with minimal modeling in terms of a
saturation scale $Q_{\text{sat}}(s)$. As a consequence, the $p_{\mathrm{T}}$
spectra should exhibit geometric scaling. In this short note we show that
recently released CMS data at $\sqrt{s}=0.9,~2.36$ and $7$~TeV fall on a
universal curve when plotted in terms of suitably defined scaling variable
$\tau$.

\end{abstract}

In Ref.\cite{1} we argued that the LHC data can be naturally described by
assuming that particles in pp high energy collisions are produced from a
saturated gluonic matter characterized by a saturation scale $Q_{\mathrm{sat}%
}^{2}(s)\sim Q_{0}^{2}(s_{0}/s)^{\lambda/2}$ where $Q_{0}\sim1$ GeV and
$s_{0}$ is the energy scale that depends on the observable in question. Power
$\lambda\sim0.2\div0.3$ is related to the growth of the gluon distribution in
proton with decreasing Bjorken $x$ as measured at HERA \cite{Stasto:2000er}. 
This behavior has a
natural explanation within the theory of saturation and the Color Glass
Condensate. For details and relevant references we refer the reader to
Ref.\cite{1}.

In Ref.\cite{1} we have argued that if the saturation momentum is the only
scale that controls $p_{\mathrm{T}}$ distributions, on dimensional grounds,
these distributions should have a geometrical scaling \cite{Stasto:2000er}
\begin{equation}
\frac{dN_{\text{ch}}}{dydp_{\text{T}}^{2}}=\frac{1}{Q_{0}^{2}}F(\tau)
\end{equation}
\newpage\begin{figure}[t]
\centering
\includegraphics[scale=0.90]{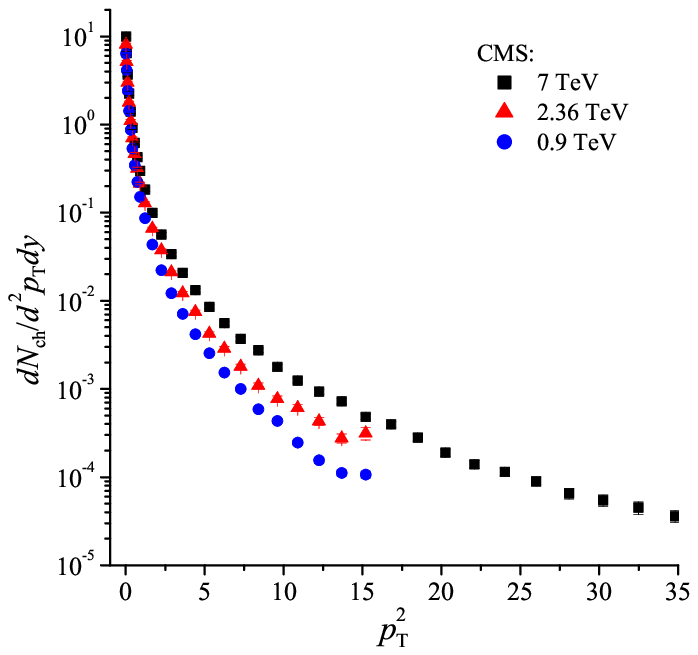}~\includegraphics[scale=0.90]{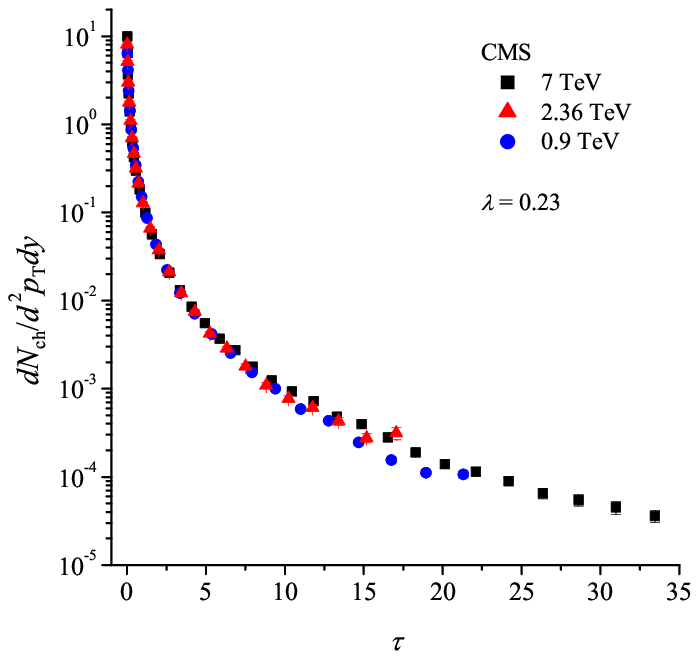}\newline%
\includegraphics[scale=0.90]{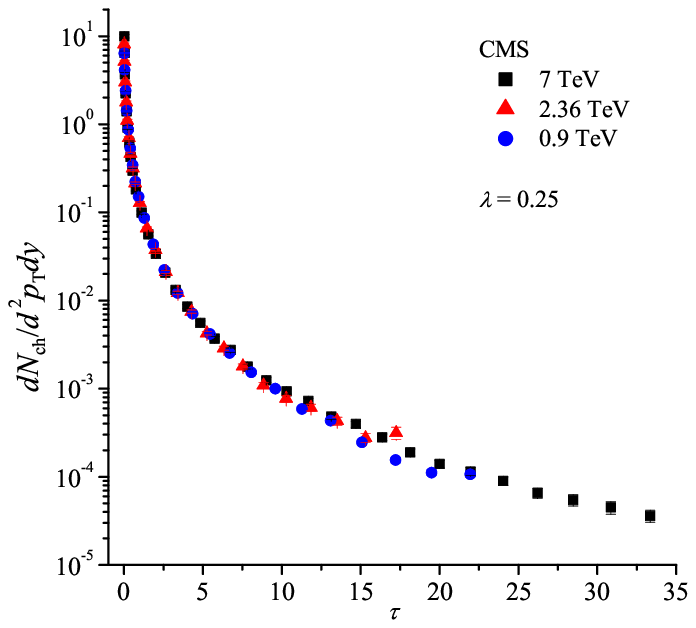}~\includegraphics[scale=0.90]{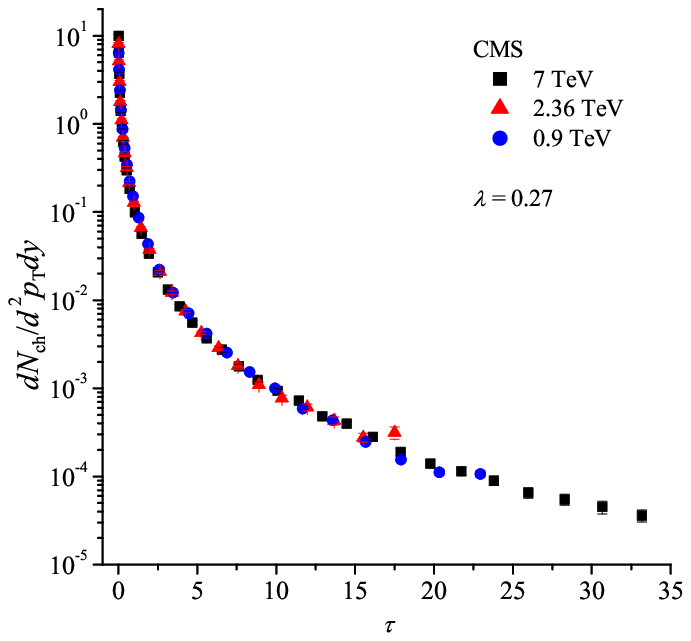}\caption{Geometric
scaling of the transverse momentum spectra that are plotted in terms of
$p_{\text{T}}^{2}$ and scaling variable $\tau$ for three choices of
$\lambda=0.23,\,0.25$ and $0.27$. }%
\label{GS}%
\end{figure}

\noindent where $Q_{0}\sim1$ GeV has been factorized for convenience. Function
$F(\tau)$ is a \textit{universal}, \textit{energy independent}, dimensionless
function of dimensionless scaling variable%
\begin{equation}
\tau=\frac{p_{\text{T}}^{2}}{Q_{0}^{2}}\left(  \frac{p_{\text{T}}}{W}\right)
^{\lambda}%
\end{equation}
where $W\sim\sqrt{s}$ (in Ref.\cite{1} $W=\sqrt{s}\times10^{-3}$ if all energy
scales are in GeV).

\begin{figure}[h]
\centering
\includegraphics[scale=0.90]{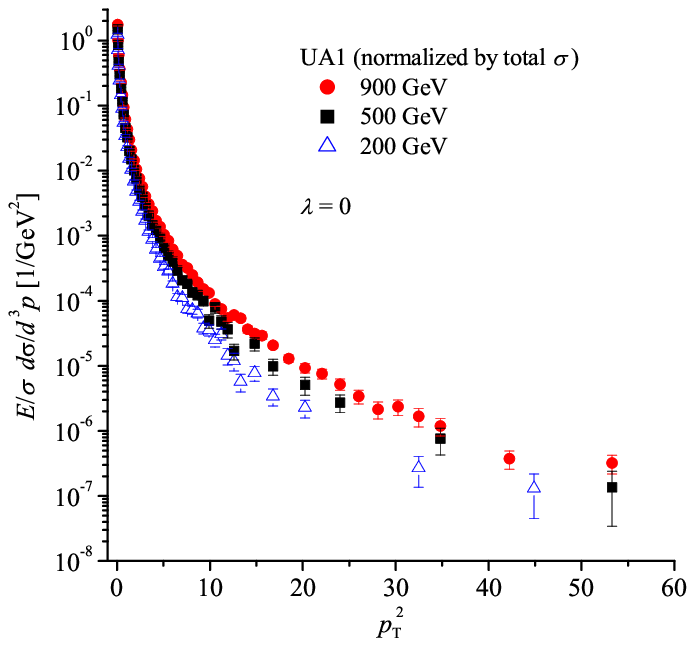}\includegraphics[scale=0.90]{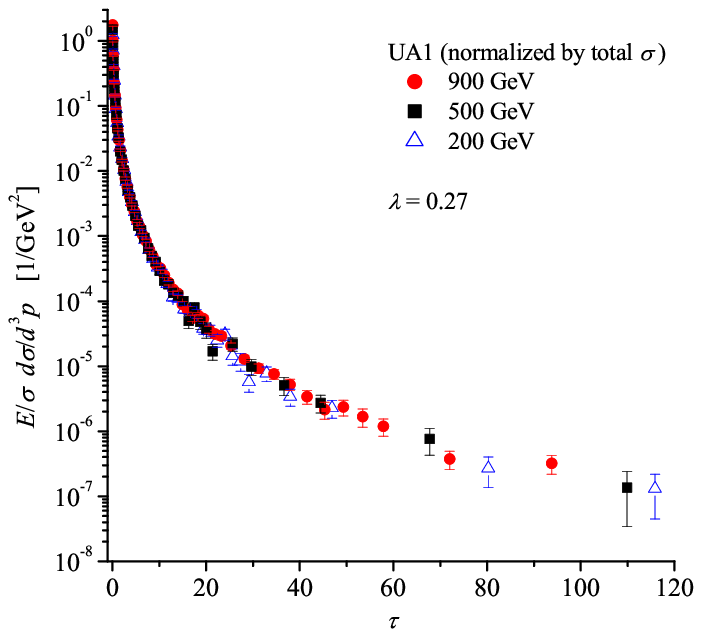}
\caption{Geometric
scaling of the transverse momentum spectra that are plotted in terms of
$p_{\text{T}}^{2}$ and scaling variable $\tau$ for $\lambda=0.27$.
Data points taken from Ref.~\cite{UA1} are normalized by total
cross-section as given in Ref.~\cite{stot}. }%
\label{GSUA1}%
\end{figure}

At the time when Ref.\cite{1} was published the $p_{\mathrm{T}}$ distributions
measured by CMS \cite{2} have not been publicly accessible yet. Therefore we
had to perform our analysis using Tsallis parametrizations of the CMS data.
Now, when $p_{\mathrm{T}}$ distributions are available from the Durham
Reaction Data Base \cite{3}, we repeat our analysis on the real data. In
Fig.\ref{GS} we plot $d^{2}N_{\text{ch}}/dyd^{2}p_{\text{T}}$ for full CMS
(pseudo)rapidity range $\left\vert \eta\right\vert <2.4$ firstly as a function
of $p_{\mathrm{T}}^{2}$ (\textit{i.e.} for $\tau(\lambda=0)$) and then as a
function of scaling variable $\tau$ for three choices of $\lambda=0.23,\,0.25$
and $0.27$. We see that indeed the data follow the universal curve exhibiting
geometric scaling. The quality of scaling depends on the actual value of
$\lambda$ and is best for the larger values shown. For all three of choices
examined in this note, the quality is very impressive. Similar conclusion can 
be drawn for the UA1 data on $\text{p}\bar{\text{p}}$
\cite{UA1} as shown in Fig.~\ref{GSUA1}.

\begin{figure}[h]
\centering
\includegraphics[scale=0.80]{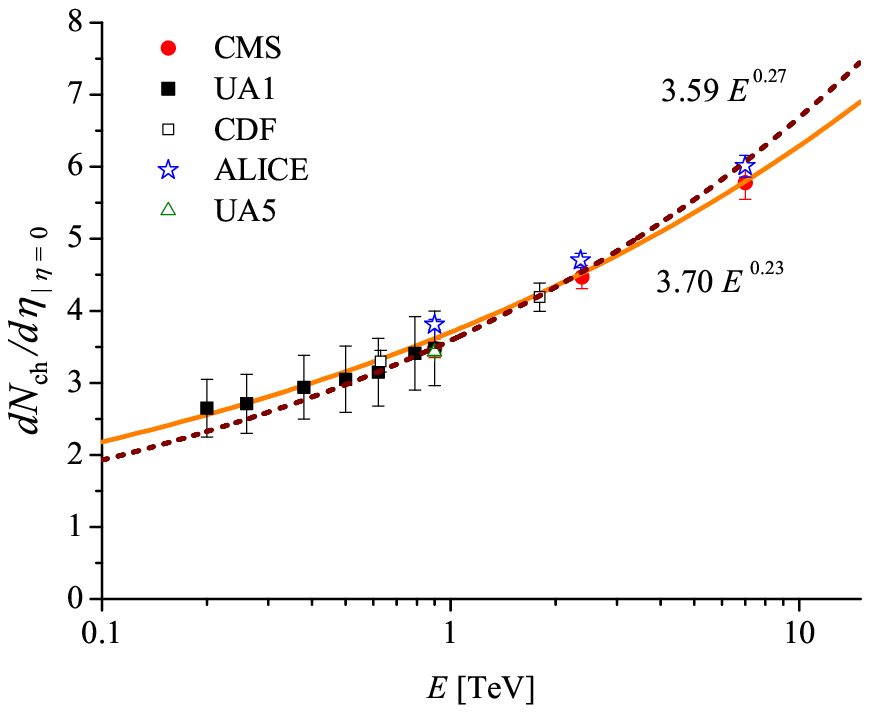}\\
\includegraphics[scale=0.80]{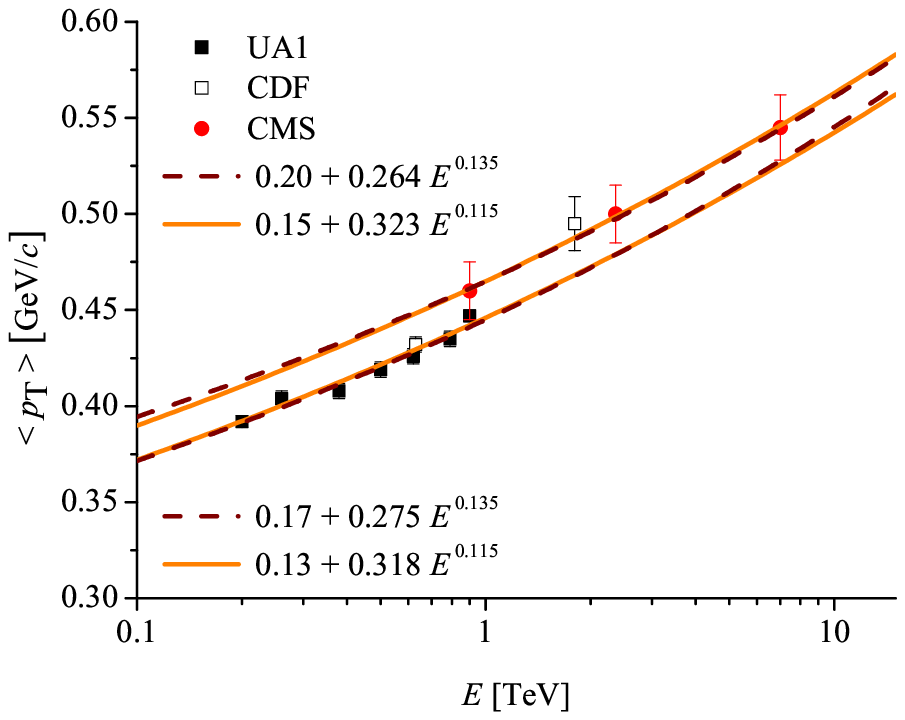}
\caption{Central multiplicity density and
average $p_{T}$ as functions of energy, plotted for energy
dependence corresponding to the two $\lambda$ values: $\lambda=0.23,\,0.27$.
Experimental data are taken from Refs. \cite{2} (CMS), \cite{UA1} (UA1),
\cite{UA5} (UA5), \cite{CDF} (CDF) and \cite{Alice} (Alice).}%
\label{distr}%
\end{figure}

We also show in Fig.~\ref{distr} the fits to the central rapidity multiplicity
and the average transverse momentum at central rapidity for the values of
$\lambda$ that describe the geometrical scaling in the transverse momentum
distributions. The data on multiplicity show a slight preference for the
smaller $\lambda$ values, but the description is quite good for all values.

This small difference between the two powers (the power $\lambda$ preferred by
geometrical scaling and $\lambda$ that fits multiplicity and $\left\langle
p_{\text{T}}\right\rangle $ growth with energy) can be easily understood as a
consequence of geometrical scaling. Indeed%
\begin{equation}
\frac{dN_{\text{ch}}}{dy}=%
{\displaystyle\int}
\frac{dp_{\text{T}}^{2}}{Q_{0}^{2}}F(\tau).
\end{equation}
Simple change of variables gives%
\begin{equation}
\frac{dp_{\text{T}}^{2}}{Q_{0}^{2}}=\frac{2}{2+\lambda}\left(  \frac{W}{Q_{0}%
}\right)  ^{\frac{2\lambda}{2+\lambda}}\tau^{-\frac{\lambda}{2+\lambda}}%
d\tau.\label{varchange}%
\end{equation}
The integral over $d\tau$ is convergent and \textit{universal}, \textit{i.e.}
it does not depend on energy. It follows from Eq.(\ref{varchange}) that the
effective power of the multiplicity growth is%
\begin{equation}
\lambda_{\text{eff}}=\frac{2\lambda}{2+\lambda}<\lambda
\end{equation}
rather than $\lambda$. For $\lambda=0.27$ we have that $\lambda_{\text{eff}%
}=0.238$.

It should be noted that in evolution equations, the parameter $\lambda$ is of order $\alpha_\text{S}$.
The deviation between $\lambda_{\text{eff}}$ and $\lambda$ is therefore of order $\alpha_\text{S}^2$.  
This gives us an estimate of the uncertainty of our considerations due to higher order effects.  
A proper treatment that builds in lograrithmic corrections and corrections due to impact parameter 
may be expected to generate deviations from our results.  It is quite exceptional that our naive 
considerations work so well.

In the meantime, after the publication of Ref. \cite{1}, other authors have
undertaken a more ambitious task to calculate particle spectra from the models
based on Color Glass Condensate that incorporate both HERA and RHIC data
\cite{4}. These studies show indeed that saturated gluonic matter dominates
particle production in $\gamma$p collisions at small Bjorken $x$ and in heavy
ion and pp collisions at high energies. If so, some universal features must be
present in the data of these three different reactions. Geometric scaling \cite{Stasto:2000er}, 
as shown in Ref. \cite{1} and in the present note, is one of them

\section*{Acknowledgments}

The research of L. McLerran is supported under DOE Contract No.
DE-AC02-98CH10886. M. Praszalowicz acknowledges support of Polish-German
PAN-DFG grant.

\end{document}